\documentclass{PoS}
\usepackage{lineno}
\usepackage{amsmath}

\title{HESS II Data Analysis with ImPACT}

\ShortTitle{HESS II Data Analysis with ImPACT}

\author{\speaker{R.D. Parsons}\\
       Max-Planck-Institut f{\"u}r Kernphysik, P.O. Box 103980, D 69029, Heidelberg, Germany\\
        E-mail: \email{daniel.parsons@mpi-hd.mpg.de}}

\author{{M. Gajdus}\\
       Institut f\"ur Physik, Humboldt-Universit\"at zu Berlin, Newtonstr. 15, D 12489 Berlin, Germany\\
        E-mail: \email{mgajdus@physik.hu-berlin.de}}

\author{{T. Murach}\\
       Institut f\"ur Physik, Humboldt-Universit\"at zu Berlin, Newtonstr. 15, D 12489 Berlin, Germany\\
        E-mail: \email{murach@physik.hu-berlin.de}}

\author{{on behalf of the H.E.S.S. Collaboration}}
\abstract{The High Energy Stereoscopic System (H.E.S.S.) very high
  energy gamma-ray telescope array has added a fifth telescope of 600
  m$^2$ mirror area to the centre of the 4 existing telescopes,
  lowering its energy threshold to the sub-100 GeV range and becoming
  the first operational IACT array using multiple telescope
  designs. In order to properly access this low-energy range however,
  some adaptation must be made to the existing event analysis.

        We present an adaptation of the high-performance
        event reconstruction algorithm, Image Pixel-wise fit for
        Atmospheric Cherenkov Telescopes (ImPACT), for performing mono
        and stereo event reconstruction with the H.E.S.S. II
        array. The reconstruction algorithm is based around the
        likelihood fitting of camera pixel amplitudes to an expected
        image template, directly generated from Monte Carlo
        simulations. This advanced reconstruction is combined with a
        multi variate analysis based background rejection scheme to
        provide a sensitive and stable analysis scheme in the sub-100
        GeV gamma-ray energy range. We will present the latest results
        of the ImPACT analysis on both simulated and real H.E.S.S. II
        data, demonstrating the behaviour of the ImPACT analysis at
        the lowest energies.}

\FullConference{The 34th International Cosmic Ray Conference,\\
		30 July- 6 August, 2015\\
		The Hague, The Netherlands}

\begin{document}
\maketitle
%\linenumbers
 
\section{Introduction}
In ground-based gamma-ray astronomy a necessary step before the
analysis of astronomical data is the reconstruction of observations
made of gamma-ray induced air showers, to determine the properties of
the initial very high energy photon. Traditionally such reconstruction
has been performed using a simple parameterisation of the roughly
elliptical gamma-ray image seen in imaging atmospheric Cherenkov
telescope (IACT) cameras, using the first and second moments of the
image. However, it has been shown that greatly
improved performance can be achieved from IACTs by performing a more
powerful maximum likelihood fit of the gamma-ray images to an expected
image template \cite{LeBohec,model++}.

Here we will present an adaptation of the high performance ImPACT
analysis \cite{ImPACT} to work with the recently commissioned
H.E.S.S. II IACT array. Such adaptation is required as an additional
telescope of $\sim$600\,m$^2$ (CT\,5) has been added to the centre of
the array, significantly reducing its energy threshold 
to below 50\,GeV. This addition creates two new classes of events,
which must be dealt with in the analysis. The \textit{mono} event class is
detected in CT\,5 only and have the lowest energy
threshold. \textit{Stereo} events on the other hand are detected in
CT\,5 and at least one other telescope, requiring two telescope
designs to be dealt with in the reconstruction process.

To demonstrate the performance of the ImPACT analysis on these
different event classes, we will present the results of two event
reconstruction types mono \& stereo. Mono
event analysis reconstructs all events using CT\,5 only (removing
information for other telescopes if present), while stereo
analyses perform reconstruction with events having images in any
two telescopes in the array.

\section{Event Reconstruction}

Before the event fit procedure can take place, first an image template
library for the two telescope types must be created for a large number
of fixed primary particle properties (energy, zenith angle, impact
distance and depth of shower maximum). These templates are created
using the same Monte-Carlo procedure described in \cite{ImPACT}, using
detailed simulation of the behaviour of each telescope type.

\subsection{Stereo Event Fitting}

Once the template library has been created event fitting can begin in
the same way as described for the H.E.S.S. I data analysis. A number
of trial primary parameters for the gamma-ray are generated from the
Hillas-parameter based analysis. At these trial positions the expected
shower image can be created by performing a multi-dimensional
interpolation of the template library and compared to the individual
camera images using the likelihood function created in
\cite{model++}. The primary particle properties can then be
reconstructed by performing a multi-dimensional fit (source position
in sky, impact position on the ground, energy and depth of shower maximum) using the
MINUIT package \cite{MINUIT}.

The adaptations to this procedure to work with stereo
events is minimal, requiring a new set of templates to be read for
CT\,5 camera images. Fitting is normally complete after 300-500
function calls, requiring around 0.5 seconds to perform the minimisation. All
stereo results presented use simple box cuts on the Mean Scaled
Width/Length parameters (described in \cite{HESScrab}). Although this
method of background rejection is rather simple it is well proven and
quite robust, in the future background rejection will be improved by the use
multivariate analysis similar to that presented in \cite{TMVA}.

\subsection{Mono Event Fitting}

\begin{figure*}
    \centering
\includegraphics[width=0.5\textwidth]{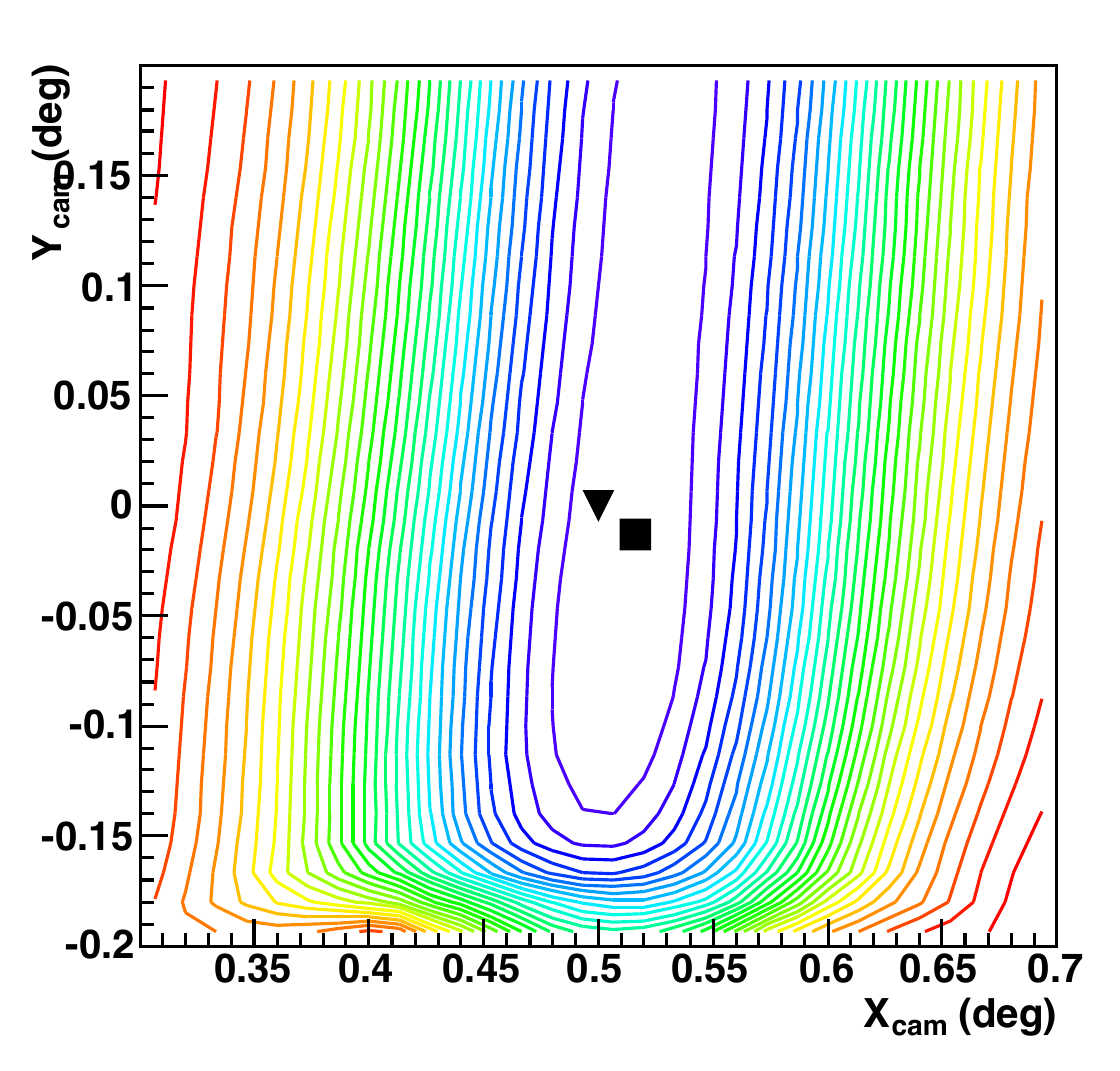}
	\caption{2D slice of the likelihood surface of a mono event,
          shown in the camera reference frame, the black triangle
          represents the simulated source position and the black
          square the ImPACT reconstructed position.}
	\label{fig-like}
\end{figure*}

In order to make the ImPACT fit converge with events with only a
single telescope image some adaptations to the fitting procedure were
required. Firstly the standard stereo event analysis was no longer
possible so a Hillas-parameter based mono telescope analysis is used
to seed the fit \cite{Murach}. In such a Hillas analysis there are two
possible solutions to the reconstructed shower direction (Head-Tail ambiguity) so both of
these starting points were tested and the highest likelihood starting
position chosen. Once this starting position is chosen the full fit
can begin, however in such a mono event analysis significantly less
information is available for fitting, leading to a far less
constrained likelihood surface (see Figure \ref{fig-like}) and
degeneracies in a number of fitted parameters. In order to avoid such
degeneracies a two-step fitting procedure was adopted, with a first
fit leaving all parameters free except X$_\mathrm{max}$ (fixed at the
expectation value for the current energy) followed by a second fit
fixing the source position and leaving all other parameters free.

Event fitting is somewhat faster than for stereo events, taking
around 0.2 seconds per event. Background rejection for mono
events is performed using a series of neural networks trained on
Hillas parameters, described in detail in \cite{Murach}. Mono
analysis cuts were optimised at 3 cut levels, \textit{safe, std \&
  loose}. These cuts are optimised with differing levels of
energy threshold (safe being highest and loose lowest) at the
expense of overall performance.

\begin{table}[b]
\begin{center}
\begin{tabular}{|c|c|c|c|c|c|c|c|c|}

\hline
Config & Amp & Npix & Local distance & $\theta^2$ (deg$^2$)& $\zeta$ &
  MSCW & MSCL\\
\hline
ImPACT mono \textbf{Safe} & 80 & 8 & 1.15$^\circ$& 0.015 & >0.9 & n/a &n/a  \\
ImPACT mono \textbf{Std} & 60 & 6 & 1.15$^\circ$&0.024 & >0.78 & n/a &n/a \\
ImPACT mono \textbf{Loose} & 40 & 4 & 1.15$^\circ$& 0.045 & >0.44 &n/a  &n/a \\
\hline
\hline
ImPACT stereo & 70, 75 & n/a & 2$^\circ$,
                                          1.15$^\circ$&0.0075 & n/a & <0.6 &
  <1.9\\
\hline

\end{tabular}
\end{center}
\caption{Image selection and background rejection cuts for the three
  H.E.S.S. mono cut configurations compared in this paper.}
\label{tab-cuts}
\end{table}%

\section{Performance}

\subsection{Simulated Data}

\begin{figure*}
    \centering
\includegraphics[width=0.49\textwidth]{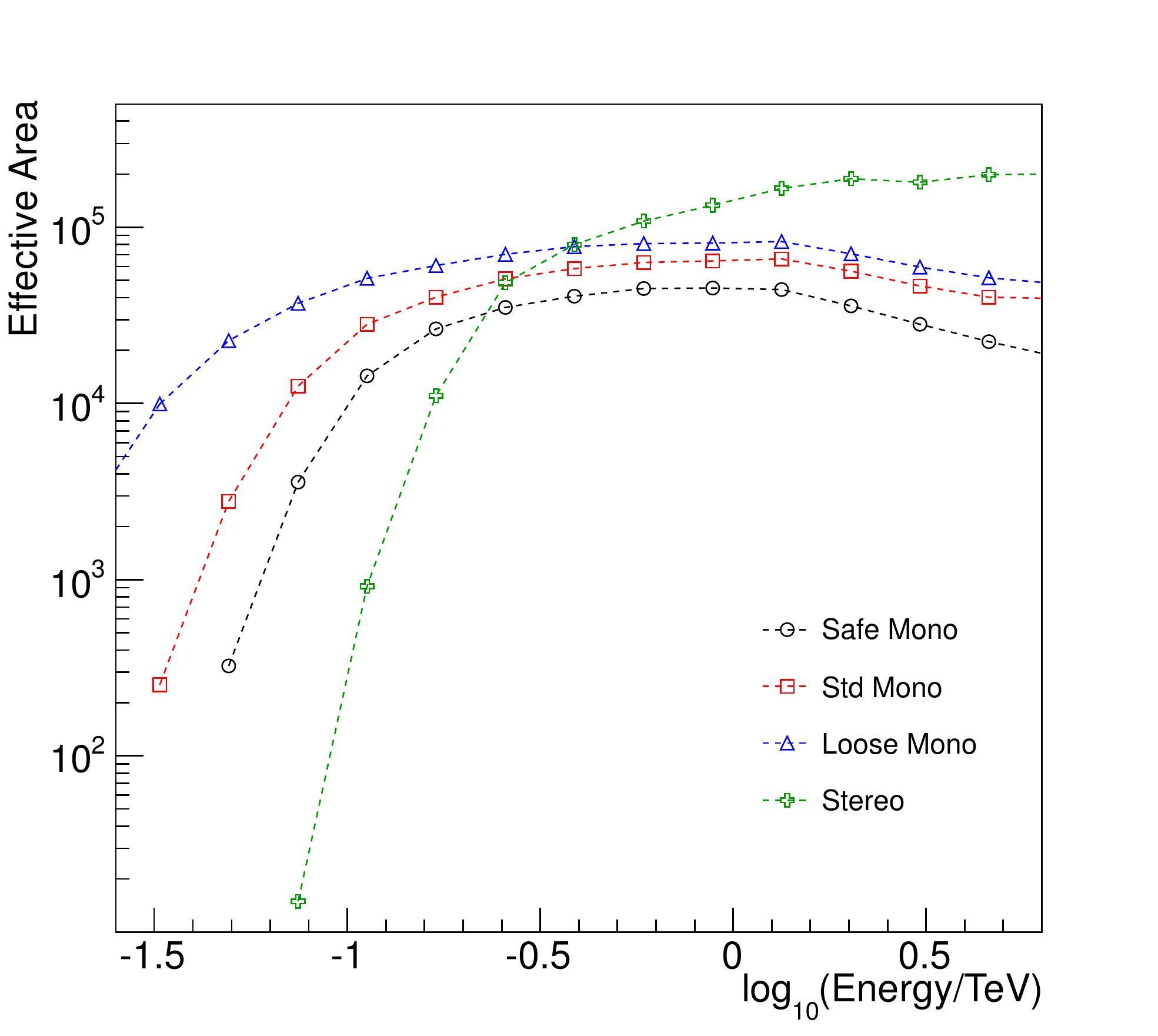}
\includegraphics[width=0.49\textwidth]{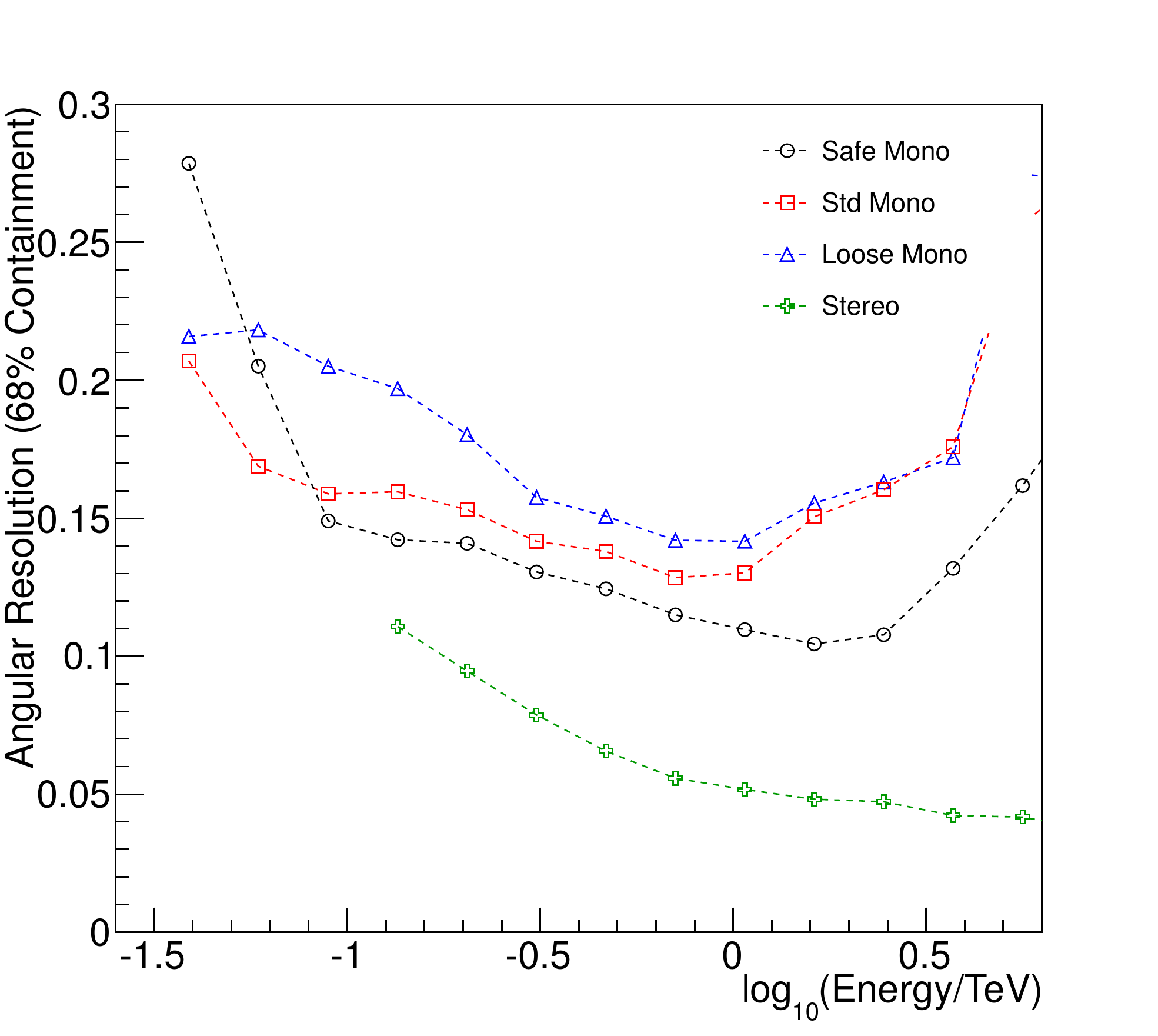}
\caption{ Angular resolution (68\% containment radius calculated from
  a double gaussian fit) and effective
  collection area as a function of the simulated gamma-ray energy, for the  of the ImPACT \textit{mono} and \textit{stereo}
  analyses for a simulated point-like source of gamma-rays at
  20$^\circ$ zenith angle. }
	\label{fig-performance}
\end{figure*}

\begin{figure*}
    \centering
\includegraphics[width=0.49\textwidth]{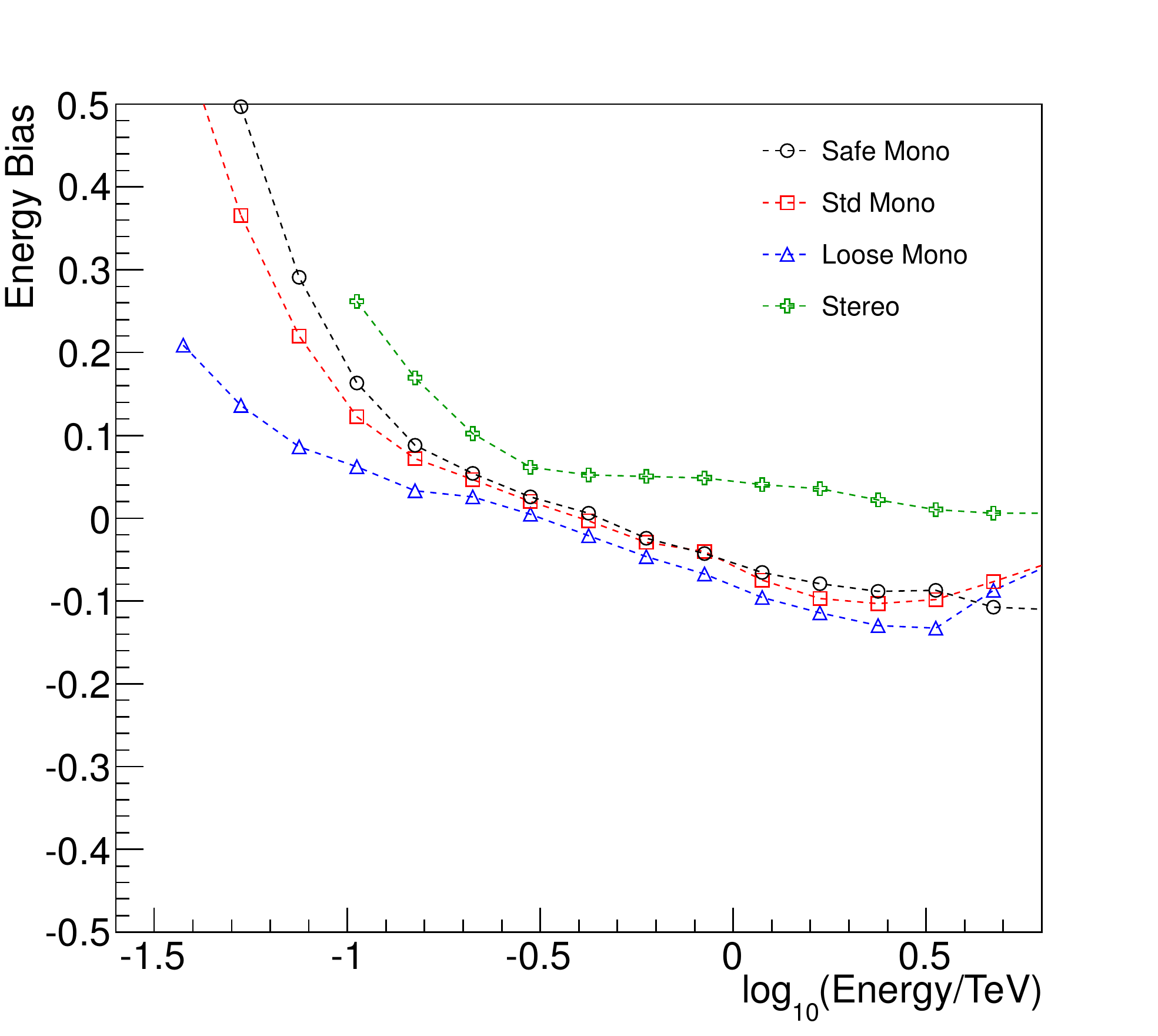}
\includegraphics[width=0.49\textwidth]{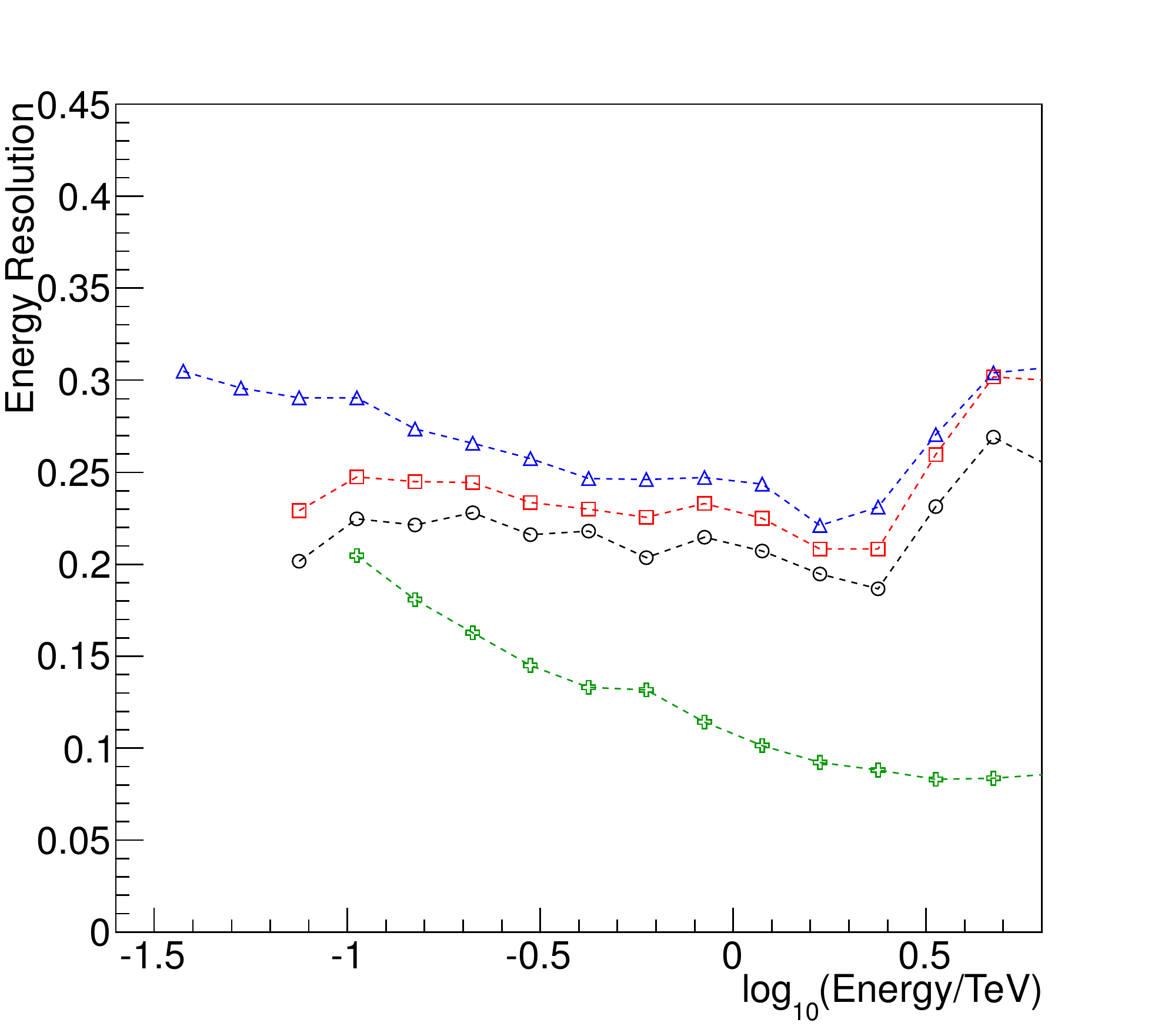}
\caption{ Energy reconstruction performance of the H.E.S.S. II ImPACT
  analyses shown by the energy bias (left) and resolution (right) as a
  function of the simulated event energy.}
	\label{fig-enperformance}
\end{figure*}

In order to demonstrate the performance of the ImPACT analysis with
H.E.S.S. II, tests were first made using simulated data. In this case
a point-like source of gamma-rays was simulated at 20$^\circ$ zenith angle. Figure
\ref{fig-performance} summarises the performance of the ImPACT
analysis for both the mono and stereo event
classes. 

Comparison of the effective areas shows that, as expected, the mono
analyses are able to retain relatively large effective areas down to
low energies (figure \ref{fig-performance}). The best low energy
performance can be achieved with the loose cuts, retaining
10$^4$\,m$^2$ of effective area down to 30 GeV. With \textit{std} and
\textit{safe} cuts the low energy area is somewhat reduced due to the
harder cuts being used. These harder cuts however, lead to significant
improvements that can be seen in the angular resolution of these
cut-sets with an angular resolution of below 0.16$^\circ$ seen for
most of the energy range, in comparison to $\sim0.2^\circ$ resolution
of the \textit{loose} cuts. For the mono analyses there is generally
an improvement in angular resolution with event energy, however
once
an energy of around 2\,TeV is reached shower images are often truncated at the
edge of the cameras reducing the reconstruction performance.  Similar
improvements can be seen in the performance of energy reconstruction
(figure \ref{fig-enperformance}). The energy biases of all 3 mono cut
sets shows some gradients, overestimating the energy of low energy
events, and underestimating at high energies. These biases are due to
the difficulties in breaking ambiguities between the impact distance
and energy for mono events.

Stereo analysis on the other hand, has quite poor effective
area at the lowest energies, due to the requirement of images in at
least 2 telescopes. Above 300\,GeV the use of the 4 H.E.S.S. I
telescopes in the analysis allows for a significantly improved
effective area. It is clear from the all performance curves that the
use of 2 telescopes in the fit procedure is extremely helpful to the
reconstruction at all but the lowest energies, resulting in angular and
energy resolutions much improved over the mono reconstruction across
the full stereo energy range. Therefore, in the majority of cases more
accurate event reconstruction can be achieved using information from
CTs 1-4 whenever available, even when only very small images are
available in these telescopes.

\begin{figure*}
    \centering
\includegraphics[width=0.5\textwidth]{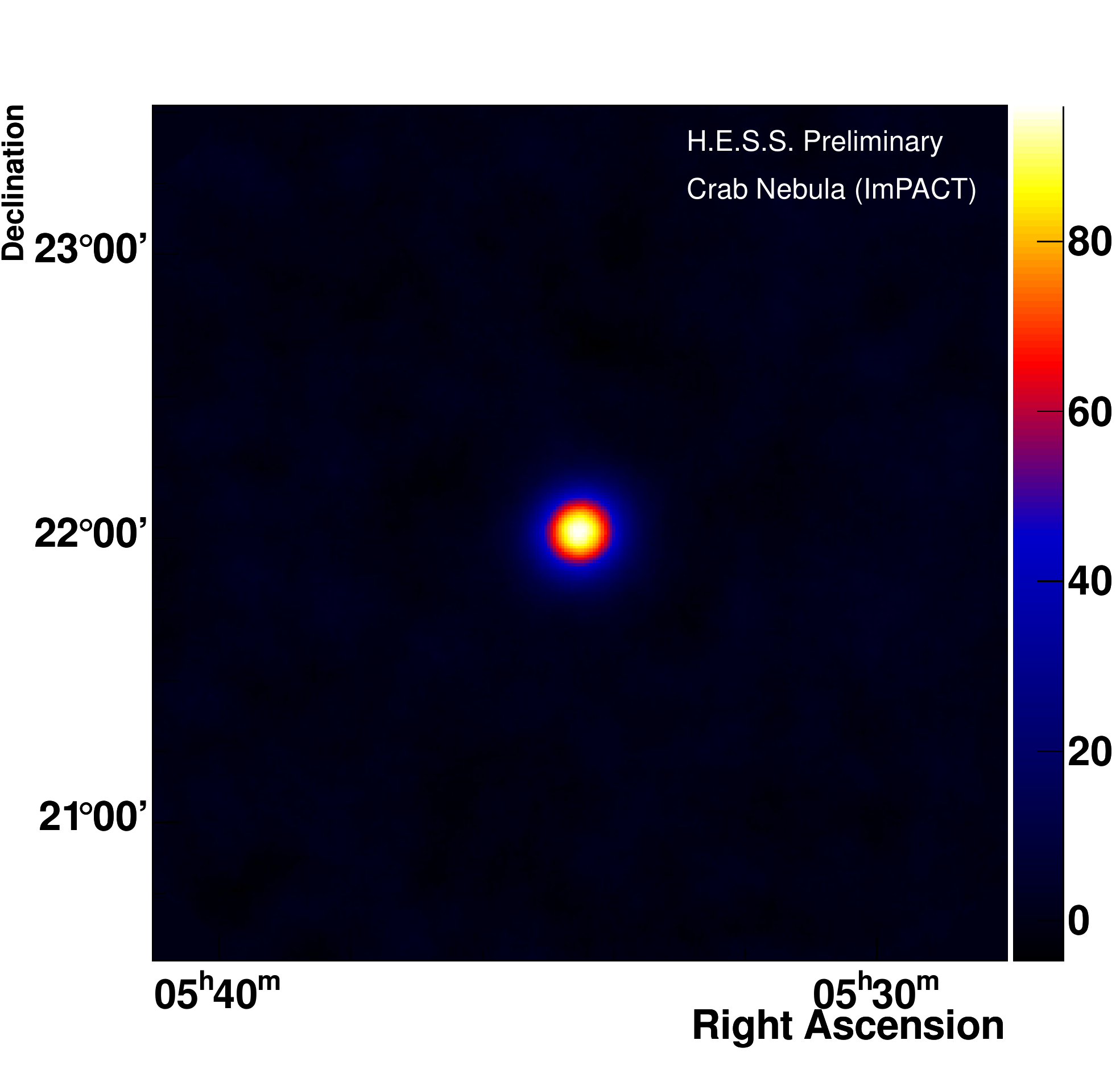}
\includegraphics[width=0.49\textwidth]{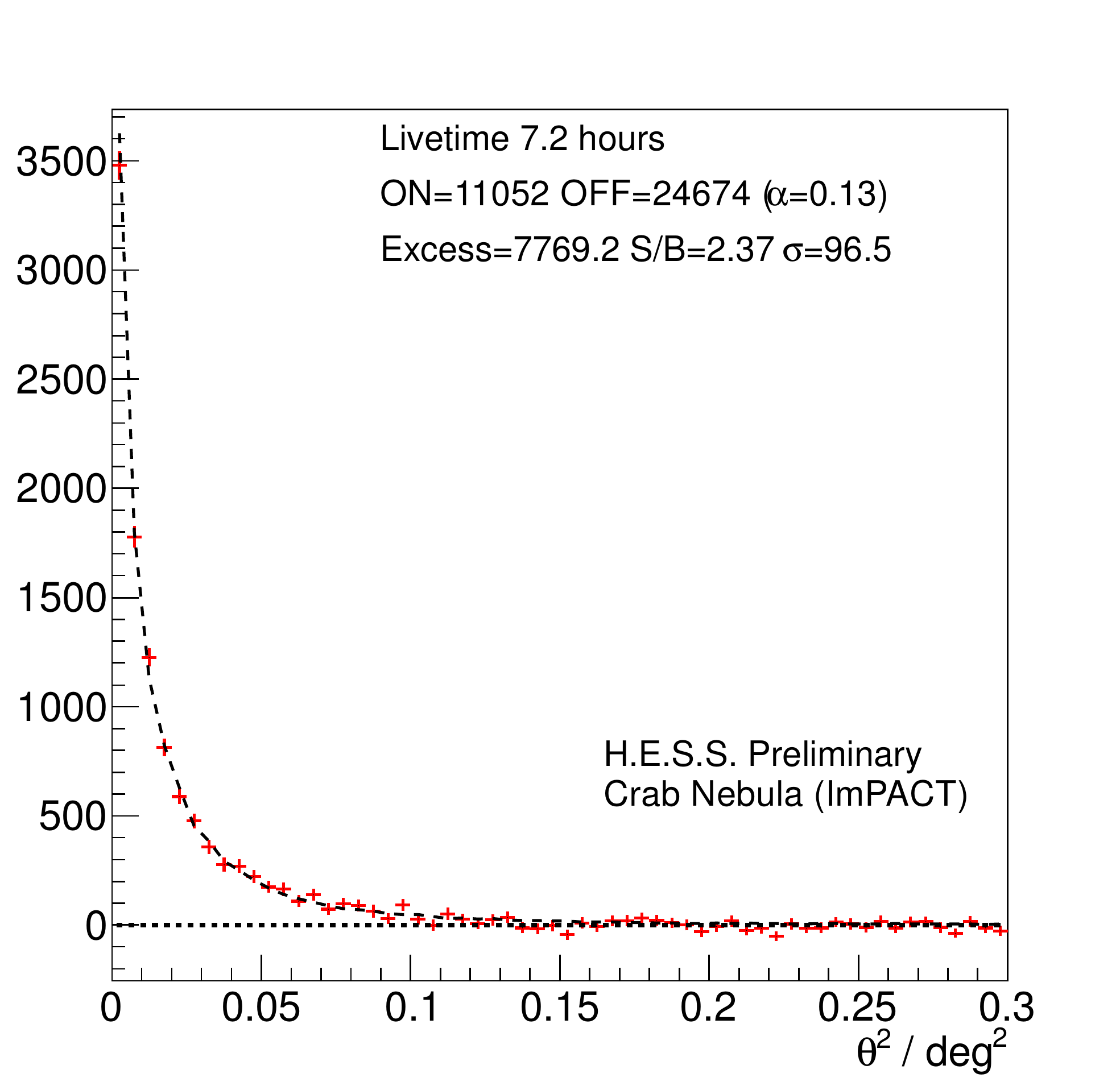}
\caption{Significance map of the Crab Nebula region using an event correlation radius of
  0.1$^\circ$ (left) and distribution of squared angular distance of
  excess events from the Crab Nebula position (right), black dashed
  lines shows the expected PSF from simulations (assuming a power-law
  spectral index of -2.4, normalised to number of counts in central 0.1 deg$^2$).}
	\label{fig-observations}
\end{figure*}

\subsection{Crab Nebula Observations}

Tests were also made using the ImPACT analysis using 7.2 hours of Crab
Nebula observations taken in late 2014. Figure \ref{fig-observations}
shows the results of this analysis using the std ImPACT mono analysis
chain. These observations were taken at an offset angle of 0.5$^\circ$
from the camera centre and a mean zenith angle of 47$^\circ$. In this
observation time the Crab Nebula was seen at a significance of more
than 96$\sigma$, with an excess of over 7700 $\gamma$-like events and
a signal-to-background ratio of 2.3. The resultant significance map is
well normalised showing a point-like source at the expected position.

\begin{figure*}
    \centering
\includegraphics[width=0.7\textwidth]{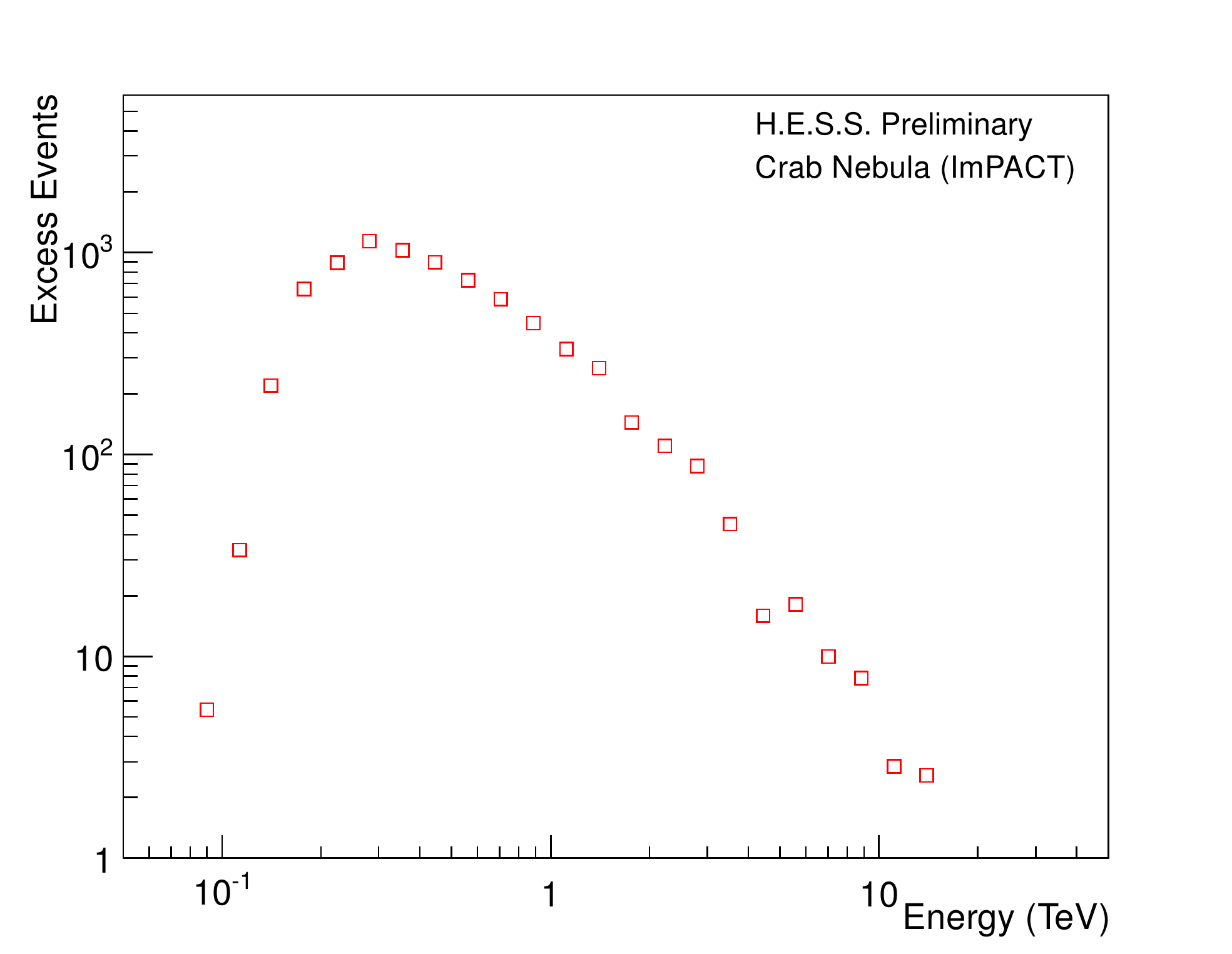}
\caption{Excess event distribution as a function of reconstructed
  event energy for 7.2 hours of Crab Nebula observations using std
  ImPACT mono cuts.}
	\label{fig-energydist}
\end{figure*}

\begin{figure*}
    \centering
\includegraphics[width=0.99\textwidth]{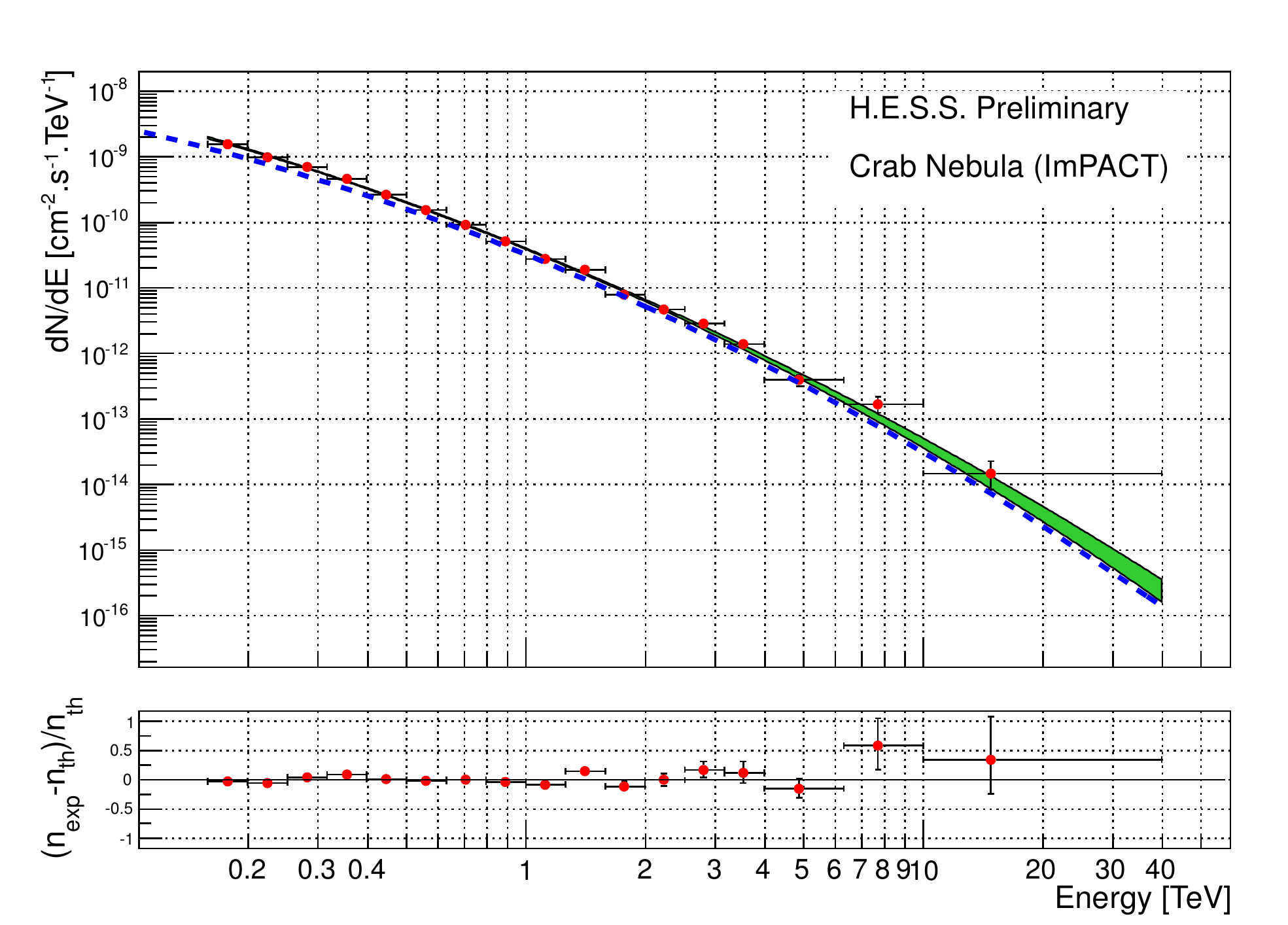}
\caption{Spectral fit of the Crab Nebula data assuming a log parabolic
model. Blue line shows the spectrum obtained by the MAGIC II telescope\cite{MAGICcrab}}.
	\label{fig-logpara}
\end{figure*}

Figure \ref{fig-energydist} shows the distribution of these excess
events as a function of reconstructed event energy. This distribution
shows excess events down to 100\,GeV (in comparison with the
300-400\,GeV energy threshold seen in \cite{HESScrab}), with the
majority of events being seen at the lowest energies
($<300$\,GeV). However, in the case of such a strong source,
significant emission is seen up to several TeV. A spectral fit was
made to this data using a log parabolic model of the form
$\phi_0 \left( \frac{E}{E_0} \right) ^ {- [\alpha + \beta
  \ln(E/E_0)]}$.
The log parabola gave an acceptable fit, finding best fit parameters
of $\phi_0=3.951\pm0.09 \times 10^{-11}$ cm$^{-2}$s$^{-1}$TeV$^{-1}$,
$\alpha=2.49\pm0.03$ and $\beta=0.203\pm0.02$, when $E_0=1$\,TeV
(Figure \ref{fig-logpara}). This spectral fit matches well
with both previously published H.E.S.S. results \cite{HESScrab} and
analyses of the same data set using other analysis chains.

\subsection{Conclusion}

We have demonstrated the adaptation of the high performance event
reconstruction algorithm ImPACT to work with the mixed telescope type
H.E.S.S. II array. This algorithm has proven both reliable and
accurate for H.E.S.S. II data analysis behaving well on both simulations
and analysis of Crab Nebula data. \\

\footnotesize{{\bf Acknowledgment:}{The support of the Namibian authorities and of the University of Namibia in facilitating the construction and operation of H.E.S.S. is gratefully acknowledged, as is the support by the German Ministry for Education and Research (BMBF), the Max Planck Society, the German Research Foundation (DFG), the French Ministry for Research, the CNRS-IN2P3 and the Astroparticle Interdisciplinary Programme of the CNRS, the U.K. Science and Technology Facilities Council (STFC), the IPNP of the Charles University, the Czech Science Foundation, the Polish Ministry of Science and Higher Education, the South African Department of Science and Technology and National Research Foundation, and by the University of Namibia. We appreciate the excellent work of the technical support staff in Berlin, Durham, Hamburg, Heidelberg, Palaiseau, Paris, Saclay, and in Namibia in the construction and operation of the equipment.}}

\end{document}